\documentclass[submission, Phys]{SciPost}

\usepackage{braket}
\usepackage{tikz}
\usepackage{dsfont}
\usepackage{siunitx, xspace}
\usepackage{array}
\usepackage{graphicx}
\usepackage{bbold}
\usepackage{subcaption}

\usepackage{amssymb}
\newcolumntype{M}[1]{>{\centering\arraybackslash}m{#1}}

\newcommand{\Hloc}{{H_{\text{loc}}}\xspace}
\newcommand{\Hbath}{{H_{\text{bath}}}\xspace}
\newcommand{\Hhyb}{{H_{\text{hyb}}}\xspace}

\newcommand{\dt}{{\Delta t}\xspace}
\newcommand{\ketpsi}{{\ket{\psi}}\xspace}

\definecolor{OliveGreen}{rgb}{0.859,0.267,0.216}
\definecolor{brickred}{rgb}{0.956,0.627,0}
\definecolor{cobalt}{rgb}{ 0.259,0.522, 0.957 }

\definecolor{bostonuniversityred}{rgb}{0.8, 0.0, 0.0}

 \usepackage{pgfplots}
  \pgfplotsset{compat=newest}

\usetikzlibrary{external}
\usetikzlibrary{calc,  shapes,shapes.gates.logic.US, backgrounds, arrows, chains, matrix, positioning, scopes, 
decorations, 
decorations.pathmorphing,patterns, fit, pgfplots.groupplots, plotmarks, }
\tikzset{>=latex}
\usetikzlibrary{external}

\newcommand{\pgfextractangle}[3]{%
    \pgfmathanglebetweenpoints{\pgfpointanchor{#2}{center}}
                              {\pgfpointanchor{#3}{center}}
    \global\let#1\pgfmathresult  
}

\usepackage{graphicx}% Include figure files
\usepackage{dcolumn}% Align table columns on decimal point
\usepackage{bm}% bold math

\usepackage{hyperref}

\begin{document}

%\preprint{APS/123-QED}
\begin{center}{\Large \textbf{Time Dependent Variational Principle for Tree Tensor Networks
}}\end{center}

\begin{center}
Daniel Bauernfeind\textsuperscript{1*},
Markus Aichhorn\textsuperscript{1},
\end{center}
\begin{center}
{\bf 1} Institute of Theoretical and Computational Physics 
  Graz University of Technology, 8010 Graz, Austria \\
* daniel.bauernfeind@tugraz.at
\end{center}

\begin{center}
\today
\end{center}

\section*{Abstract}
{\bf
We present a generalization of the Time Dependent Variational Principle (TDVP) to any finite sized loop-free tensor 
network. The major advantage of TDVP is that it can be employed as long as a representation of the Hamiltonian in the 
same tensor network structure that encodes the state is available. Often, such a representation can 
be found also for long-range terms in the Hamiltonian. As an application we use TDVP for the Fork Tensor 
Product States tensor network for multi-orbital Anderson impurity models. We demonstrate that TDVP allows to account 
for off-diagonal hybridizations in the bath which are relevant when spin-orbit coupling effects are important, or when 
distortions of the crystal lattice are present. }

\section{Introduction}
The development of the Density Matrix Renormalization Group (DMRG)~\cite{WhiteDMRG,SchollwoeckDMRG_MPS} was an immensely 
important milestone in our understanding of one-dimensional quantum systems. The subsequent realizations that DMRG 
produces Matrix Product States~\cite{Ostlund_DMRGprodMPS} (MPS) and that it can be formulated as a variational 
method~\cite{Dukelsky_RelationDMRGMPS}, ultimately led to the development of numerous approaches using not only MPS 
but also general Tensor Networks to handle quantum systems. Notable examples are the Projected Entangled Pair States 
(PEPS)~\cite{VerstraeteCirac_PEPS_2004,MurgVerstraete_PEPS2_2007}, the Multi-scale Entanglement Renormalization Ansatz 
(MERA)~\cite{VidalMera} and so-called Tree-Tensor Networks 
(TTN)~\cite{Delgado_DMRGDendrimers_2002,Depenbrock_BetheLatticeImagTEBD_2013,Otsuka_DMRGTTN_1996, 
Friedman_DMRGCayleyTree_1997,Gerster_UnconstrainedTree_2014,SilviCirac_binaryTree_2010, 
MurgVerstraete_TTN_2010,KlaasVerstraete_TTNChemistryDMRG_2018,TagliacozzoVidal_2dTTN_2009,ShiVidal_TTN_2006} including 
also the recently developed Fork Tensor Product States (FTPS) method~\cite{FTPS,THESIS}.
\\
Among the most important properties of tensor networks is whether their graph is loop-free, i.e., whether there exists 
only a single path from one tensor to any other. While PEPS and MERA are not loop-free, the TTNs and MPS are. 
Cutting any edge of a loop-free network, results in two separated segments and therefore gives a notion of left and 
right with respect to this edge. This in turn allows a controlled truncation scheme based on the Schmidt-decomposition 
of quantum states in the spirit of DMRG.
\\
One of the major reasons behind the success of tensor networks are the celebrated area laws of 
entanglement~\cite{Eisert_AreaLaw} stating that the entanglement of ground states of gapped Hamiltonians with short-range couplings is 
proportional to the surface area connecting the two regions. MPS in 1-d and PEPS in 2-d efficiently encode quantum 
states obeying these area laws and are hence efficient parametrizations. In addition, MPS-based time evolution for one 
dimensional systems is an important method to calculate dynamical 
properties~\cite{Eisert_NonEq_2015,Matthias_MagnetizationProfiles_2019,DanielViktor_FrontDynamics, 
Collura_domainWallexsolution_2018}. Approaches to perform the real-time evolution include, among others, the 
Time-dependent Density Matrix Renormalization Group (tDMRG)~\cite{DaleySchollwoek_tDMRG,White_tDMRG}, the closely 
related Time Evolving Block Decimation (TEBD)~\cite{VidalTEBD1,VidalTEBD2} as well as the Time Dependent Variational 
Principle (TDVP)~\cite{HaegemanTDVP1,HaegemanTDVP2,Lubich_TDVP}. An in depth comparison of several time evolution algorithms 
performed in Ref.~\cite{Hubig_CompareTevoAlgs} came to the conclusion that while all approaches have strengths 
and weaknesses, TDVP is among the most reliable methods to perform the time evolution.
\\
While time evolution approaches for MPS are well established, much less has been done for general tensor networks. 
So far, mostly TEBD (and variations) have been used, for example for the MERA network~\cite{Rizzi_MERATevo_2008}, 
for PEPS~\cite{PhienVidal_PEPSTEBD_2015,Hubig_IPEPS_RealTimeTEBD,Czarnik_IPEPSTEBD} and for 
TTNs~\cite{Depenbrock_BetheLatticeImagTEBD_2013,LiXiang_BetheInfiniteDMRG_2012,
Nagaj_TEBDImagTevoTTN_2008,ShiVidal_TTN_2006,FTPS}. The advantage of TEBD is its relative simplicity, since it 
effectively boils down to a repeated application of short range operators obtained from a Suzuki-Trotter 
decomposition~\cite{SuzukiDecomp} of the full time-evolution operator. 
\\
However, one of the major disadvantages of TEBD is that it can become difficult to 
implement for more complicated Hamiltonians, especially when long-range couplings are present. One approach to treat such
couplings is an MPO-based approach introduced by Zaletel et al.~\cite{Zaletel_longrange} in which a MPO approximation of the time-evolution 
operator is constructed. Alternatively, TDVP circumvents this problem by only demanding a Hamiltonian represented in the same tensor network structure as the state which is often easy to find. Additionally, TDVP in its single-site variant exactly respects conserved quantities of the 
Hamiltonian like energy or magnetization~\cite{HaegemanTDVP2}. Although some works applied TDVP to more general 
tensor networks~\cite{RamsEntBarrier,schroederF_TDVP}, it is not obvious how these algorithms work in detail and how it 
can be generalized. A notable exception is Ref.~\cite{Lubich_TTNTDVP} which introduces TDVP for binary TTNs. Parallel to these developments in the tensor network physics community, very similar approaches to TDVP have been developed in quantum chemistry under the name of Multi-layer Multi-Configurational Time-Dependent Hartree approach~\cite{Meyer_MCHF1,Meyer_MCHF2,Manthe_MCHF,Manthe_MCHF2}. These methods effectively generate tensor networks by repeatedly grouping degrees of freedom together and transforming them with (time dependent) basis transformations into new degrees of freedom.
\\
A more practical motivation for the formulation of TDVP for TTNs are Dynamical Mean-Field Theory (DMFT) 
calculations using the FTPS tensor network. So far, this approach has been used for so-called diagonal 
hybridizations only. On the other hand, real materials often exhibit off-diagonal hybridizations, 
which can for example come from spin-orbit coupling, or from distortions of the crystal lattice. For off-diagonal 
hybridizations, the TEBD approach used so far~\cite{FTPS,THESIS} is difficult to generalize and we hence choose to use 
TDVP in these situations.
\\
Although part of the motivation for this work comes from the FTPS tensor network, in this paper we formulate TDVP for 
\emph{general} loop-free and finite-size tensor networks. After establishing the relevant concepts 
of TTNs in Sec.~\ref{sec:TTNS}, we generalize TDVP to these networks in Sec.~\ref{sec:TDVP_TTN}. Finally in 
Sec.~\ref{sec:TDVP_FTPS} we show how this approach can be used for the FTPS tensor network and that it can be applied to 
off-diagonal hybridizations.

\section{Tree Tensor Networks Basics}\label{sec:TTNS}
\begin{figure}
 \centering
\includegraphics{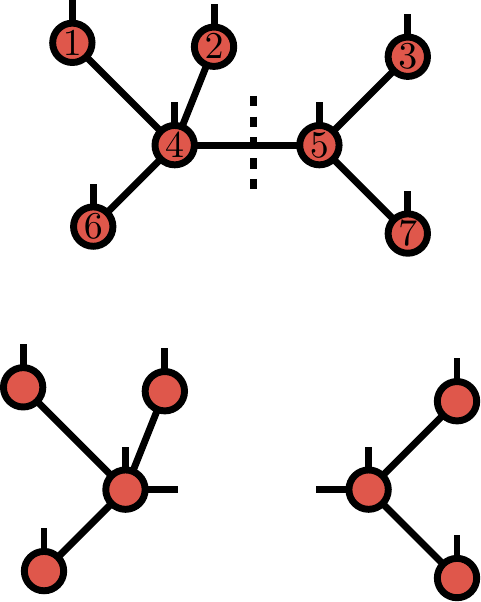}
\caption{ Example of a TTN with 7 tensors with different numbers of link-indices on each site. Each dot represents a 
tensor and each line an index, where a connected line implies summation over this index. The open lines 
are the 
physical indices $s_1 \cdots s_N$ ($N=7$), while the connected lines are the link indices $q_1 \cdots q_L$ ($L=6$). 
Cutting the link between sites $4$ and $5$, as indicated by the dashed line, results in two disconnected tensor network 
segments and defines a notion of left and right at each link. In this example, sites $1$, $2$, $3$, $6$ and $7$ are the 
leaves of the TTN. }
\label{fig:TTN}
\end{figure}

In this section, we discuss concepts of TTNs relevant for the formulation of TDVP. All these properties 
are generalizations of the corresponding concepts for MPS. Although these have been discussed previously in 
several publications (see for example 
Refs.~\cite{ShiVidal_TTN_2006,Silvi_TensorNetworksAnthology_2019,MurgVerstraete_TTN_2010}), here, we present them 
in a 
format that will suit us for the subsequent formulation of the TDVP algorithm.
\subsection{TTNs}x
Any state $\ketpsi$ of a quantum system consisting of $N$ sites with local basis states $\ket{s_i}$ on site $i$ 
can be expanded in the corresponding product basis:
\begin{equation}
 \ketpsi = \sum_{s_1 \cdots s_N} c_{s_1 \cdots s_N} \ket{s_1 \cdots s_N}.
\end{equation}
The coefficient $c_{s_1 \cdots s_N}$ is interpreted as a rank-N tensor with indices $s_1 \cdots s_N$. Tensor networks 
represent this rank-N tensor as a product over tensors of much smaller rank:
\begin{align}\label{eq:TTN_definition}
 c_{s_1 \cdots s_N} = \sum_{ q_1 \cdots  q_L  } T^{s_1}_{ Q_1 } \cdot T^{s_2}_{Q_2} \cdots T^{s_N}_{ Q_N} \nonumber \\
 \ket{\psi[T]} = \sum_{\stackrel{s_1 \cdots s_N}{q_1 \cdots  q_L}} T^{s_1}_{ Q_1 } \cdot T^{s_2}_{Q_2} \cdots T^{s_N}_{ 
Q_N} \ket{s_1 \cdots s_N}.
\end{align}
Each tensor $T^{s_i}_{ Q_i } \equiv T^{s_i}_{ q_1 q_2 \cdots q_{r_i} }$ has a set of auxiliary indices $Q_i =\{ q_k : 
q_k \text{ is attached to node $i$ } \}$ such that 
each auxiliary index is part of exactly two tensors. We call $r_i = |Q_i|$ the number of indices of the tensor on site 
$i$. Additionally, we attached to each tensor a physical index as for example in the FTPS tensor network. While for 
general TTNs not all tensors have a physical index, the following results can be straightforwardly generalized by just 
removing the physical index from the notation. Alternatively, every tensor without a physical index could be 
interpreted 
as having a dummy index with just a single entry corresponding to a single state, say $\ket{0}$, onto which the 
Hamiltonian 
acts as an identity $H 
\ket{0} = \ket{0}$. Note that if all sites have a physical index, the number of links is $L=N-1$. 
In the following, we will often omit sums over auxiliary indices $\sum_{ q_1 \cdots  q_L  }$ and assume Einstein 
convention for the summations.
\begin{figure}
 \centering
\includegraphics{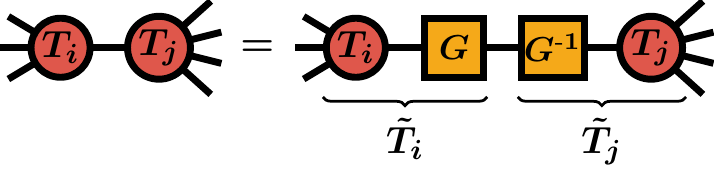}
\caption{ Gauge degree of freedom in tensor networks. At each link, one can insert an identity $\mathbb{1} = G \cdot 
G^{-1}$ without changing the physical state $\ketpsi$. By absorbing $G$ into one tensor and $G^{-1}$ into the other, we 
obtain a different representation of the same state $\ketpsi$. }
\label{fig:TensorGauge}
\end{figure}

\begin{figure}
 \centering
\includegraphics{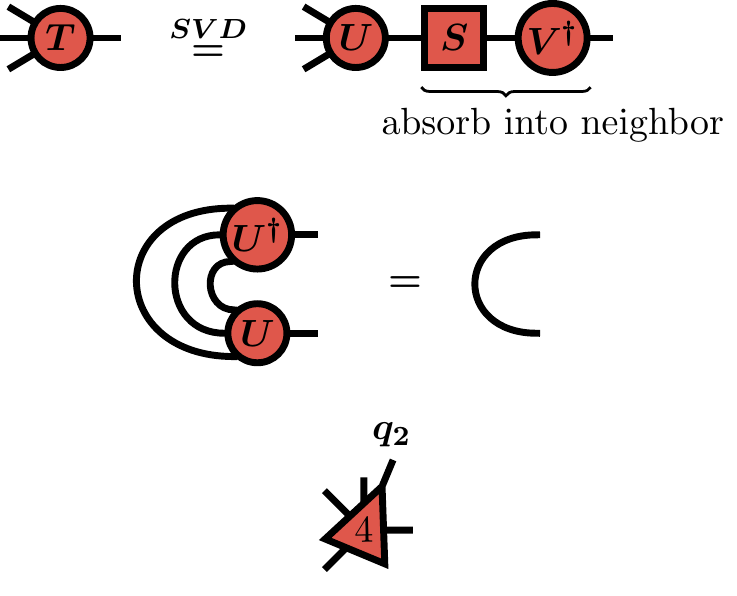}
\caption{With an SVD, we can orthogonalize a tensor towards one of its neighbors with which it shares an index. 
\emph{Top: } Tensor $T$ is reshaped into a matrix and the $U$-matrix of its SVD is used as new tensor. 
$S\cdot V^\dag$ is absorbed into the neighboring tensor. \emph{Middle:} Graphical representation of $U^\dag \cdot U = 
\mathbb{1}$. \emph{Bottom:} A tensor that is normalized towards one of its neighbors is depicted as a 
triangle pointing in the direction of this neighbor. The picture shows tensor 4 of the TTN in Fig.~\ref{fig:TTN} 
orthogonalized towards tensor $2$. Let us call the index connecting tensor $4$ with tensor $2$, $q_2$. In this case, we 
denote tensor $4$ as $ \left( T^{\mathcal{N}[q_2]} \right)^{s_4}_{Q_4}$. }
\label{fig:SVD}
\end{figure}
An example for a TTN with $N=7$ sites and $L=6$ auxiliary indices (links) is shown in Fig.~\ref{fig:TTN}. The property 
distinguishing a TTN from a general tensor network is that the graph of a TTN  is \emph{loop-free}, i.e., to move from 
one site to any other there is only one unique path along the links. This also implies that by cutting any link, the 
tensor networks splits into two disconnected segments. Therefore, at each link there is a notion of \emph{left} and 
\emph{right} which is a first hint towards the capability of TTNs to access the Schmidt decomposition and 
with it also the reduced density matrix as demonstrated below. We also define the \emph{leaves} of the TTN as 
all tensors with just a single link index. For convenience, we assume site $N$ to be a leave of the TTN.
Since TTNs are \emph{loop-free}, one can also define a measure of distance $d_{ij}$ between two sites $i$ and $j$ 
given by the number of links one has to traverse to move from site $i$ to site $j$.

\subsection{Tensor Gauge and Orthogonality Center}
The representation of a quantum state as a tensor network is highly non-unique. This \emph{gauge degree of freedom} can 
be used to obtain useful representations of the same quantum state as a TTN with certain properties, which can speed up 
calculations dramatically. As shown in Fig.~\ref{fig:TensorGauge}, at each link one can insert an identity $\mathbb{1} = 
G \cdot G^{-1}$ for any invertible matrix $G$. By absorbing $G$ into one tensor and $G^{-1}$ into the other, a 
different 
representation of the same state is reached. In this part, we make use of this gauge degree of freedom to define an 
orthogonality 
center of the TTN. 
\\
A tensor $T^{s_i}_{Q_i}$ can be orthogonalized towards one of its neighbors with which it shares link $q_k$ 
%using a QR decomposition 
as follows:
\begin{itemize}
 \item Reshape $T^{s_i}_{Q_i}$ into a matrix $T_{ (s_i, Q_i \setminus q_k), (q_k) }$ with rows $(s_i, Q_i \setminus 
q_k)$ and column $(q_k)$.
\item Perform an SVD (a QR decomposition is faster): $T_{ (s_i, Q_i \setminus q_k), (q_k) } = \sum_\alpha U_{ (s_i, Q_i \setminus q_k), (\alpha) } \cdot S_{\alpha} \cdot V^\dag_{(\alpha),(q_k)}$
\item Keep $U_{ (s_i, Q_i \setminus q_k), (\alpha) }$ as the new local tensor on site $i$ and absorb $S\cdot V^\dag$ 
into the corresponding neighbor by multiplying $S\cdot V^\dag$ onto it (formally also relabel $\alpha \to q_k$).
\end{itemize}
The SVD as well as the QR decomposition guarantees that the new site tensor has the property (see Fig.~\ref{fig:SVD})
\begin{equation*}
 \left( U^\dag \cdot U \right )_{(\alpha),(\alpha')} = 
\sum_{s_i, Q_i \setminus q_k } (U^\dag)_{(\alpha),(s_i, Q_i \setminus q_k)} U_{(s_i, Q_i \setminus q_k),(\alpha')} = 
\delta_{(\alpha),(\alpha')}.
\end{equation*}
 For tensors orthogonalized towards their neighbor 
along link $q_k$ we introduce the notation $\left( T^{\mathcal{N}[q_k]} \right )^{s_i}_{Q_i}$ (see Fig.~\ref{fig:SVD} 
bottom).
\\
As already mentioned, in TTNs there is a unique path between any two tensors. Therefore, by orthogonalizing a tensor 
towards one of its neighbors, we also orthogonalize it towards all other tensors, which can be reached via this 
neighbor. For example, to orthogonalize tensors $1$, $2$ and $6$ in Fig.~\ref{fig:TTN} towards tensor $3$, we 
orthogonalize all of them towards tensor $4$ with the procedure described 
above. 
\\ 
Next, let us introduce orthogonality centers. Site $i$ is an orthogonality center with tensor $C^{s_i}_{Q_i}$ if all 
tensors of all other sites are orthogonalized towards site $i$. To obtain such an orthogonality center, we can use 
the following algorithm:
\begin{enumerate}
  \item Find the maximum distance $d_{max}$ between site $i$ and any other site in the TTN.
  \item Initialize $d = d_{max}$ and perform the following steps until $d=0$
  \begin{itemize}

	  \item Orthogonalize all sites $j$ that are at distance $d$ from site $i$ towards site $i$, i.e., towards the 
single neighbor on the path from $j$ to $i$.
	  \item Reduce $d$ by one $d \to d-1$.
 \end{itemize}
\end{enumerate}

For example, to orthogonalize the TTN shown in Fig.~\ref{fig:TTN} towards site $4$, we first orthogonalize sites 
$3$ and $7$ towards site $5$ and then sites $1$, $2$, $6$ and $5$ towards site $4$. 
\\ 
\begin{figure}
 \centering

 \parbox{8cm}{
\includegraphics{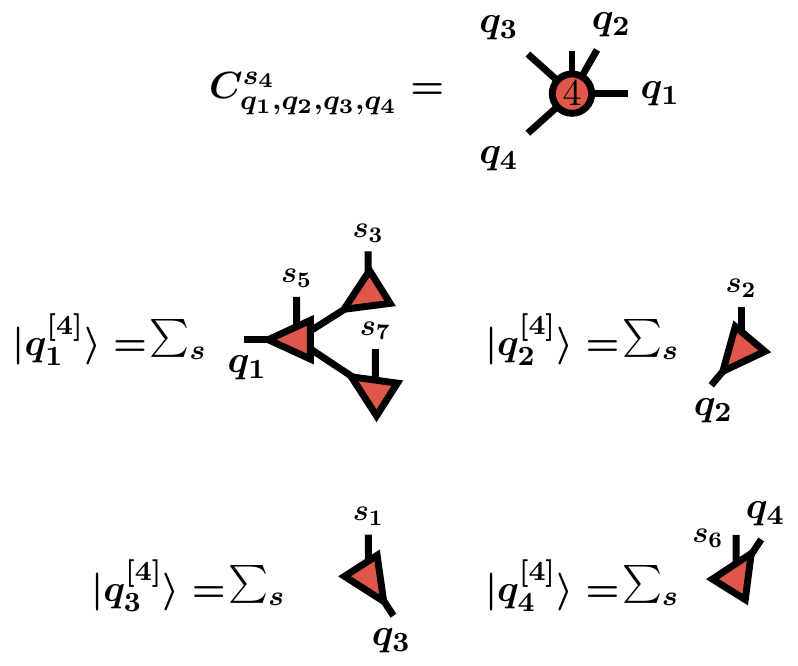} }
\qquad
\begin{minipage}{3cm}
\includegraphics[scale=0.7]{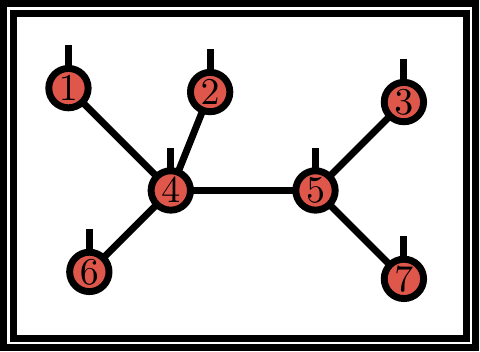}
\end{minipage}

\caption{If the orthogonality center of the TTN depicted in Fig.~\ref{fig:TTN} is placed on site $4$, the center 
tensor $C$ has four link indices $q_1 \cdots q_4$. Each of these links corresponds to one of four mutually orthogonal 
set of states $\ket{q^{[4]}_1} \cdots \ket{q^{[4]}_4}$. This orthogonality is a direct result of the orthogonality 
property of the $U$-matrices of the SVD used on all sites except site 4 (see also Fig.~\ref{fig:SVD}). The inset is a reminder of the TTN used in this section. }
\label{fig:orthoCenter}
\end{figure}

The wave function of a TTN with orthogonality center $C^{s_i}_{Q_i}$ can be written as:
\begin{align}
 \ketpsi &= \sum_{ \stackrel{q_1, q_2, \cdots, q_{r_i} \in Q_i}{s_i}} C^{s_i}_{q_1,q_2,\cdots,q_{r_i}} 
\ket{s_i} \ket{q^{[i]}_1} \ket{q^{[i]}_2} \cdots \ket{q^{[i]}_{r_i}} \nonumber \\
\ket{q^{[i]}_k} &=  \sum_{ s_1 \cdots s_r \in S^i_{q_k} } \left( T^{s_1}_{Q_1} \cdot T^{s_2}_{Q_2} \cdots T^{s_r}_{Q_r}
\right )_{q_k} \ket{s_1 \cdots s_r}\nonumber \\
\langle q^{[i]}_k | q^{[i]'}_k \rangle &= \delta_{q_k, q_k'}.
\end{align}
Here, $S^i_{q_k}$ is the segment of the tensor network that is obtained by cutting index $q_k$ and which does not 
contain site $i$. The states $\ket{q^{[i]}_{k}}$ form an orthogonal basis and are defined in Fig.~\ref{fig:orthoCenter} 
for the TTN of Fig.~\ref{fig:TTN} with orthogonality center on site $i=4$.
\\
Orthogonality centers allow to easily calculate local observables acting on the orthogonality center. For example, the expectation value of the operator $\hat{A} = \sum_{s_i,s_i'} A^{s_i', s_i} \ket{s_i'}\bra{s_i}$ acting non-trivially only on site $i$, reduces to:
\begin{equation}
\langle \psi | \hat{A} | \psi \rangle = \sum_{ \substack{q_1 q_2 \cdots q_{r_i} \in Q_i \\ s_i,s_i'} 
} \bar{C}^{s_i'}_{q_1 q_2\cdots q_{r_i}} \cdot A^{s_i' s_i} \cdot C^{s_i}_{q_1q_2,\cdots q_{r_i}},
\end{equation}
where the bar denotes complex conjugation. Orthogonality centers hence reduce the costly contraction over the whole 
tensor network, to a simple contraction over the center tensor $C$ only.

\subsection{Truncation of TTNs}
A TTN with an orthogonality center allows to calculate any Schmidt decomposition of the quantum state with 
respect to the two parts of the system defined by cutting any of the links of the orthogonality center.
To do so, we reshape the center tensor $C$ into a matrix with physical index $s_i$ and one of the 
links $q_k$, combined into the row index and all other indices into the column indices, i.e., 
$C^{s_i}_{q_1,q_2,\cdots,q_{r_i}} = C_{(s_i,q_k),(Q_i \setminus q_k)}$. The Schmidt decomposition then follows from an 
SVD of this matrix:
%\begin{gather}
\begin{align}
 &C_{(s_i q_k),(Q_i \setminus q_k)} = \sum_\alpha U_{(s_i q_k),(\alpha)} \cdot S_\alpha \cdot
(V)^\dag_{(\alpha),(Q_i \setminus q_k)} \nonumber \\
\Rightarrow \ketpsi &= \sum_\alpha S_\alpha \underbrace{\sum_{s_i q_k} U_{(s_i q_k),(\alpha)} \ket{q^{[i]}_k} 
\ket{s_i}}_{\ket{L}_\alpha}  \cdot \underbrace{\sum_{Q_i \setminus q_k} (V)^\dag_{(\alpha),(Q_i \setminus q_k)} \bigotimes_{l \neq k} 
\ket{q^{[i]}_l} }_{\ket{R}_\alpha} \nonumber \\
 &= \sum_{\alpha} S_\alpha \ket{L}_\alpha \ket{R}_\alpha. \label{eq:schmidtTTN}
\end{align} %\\ ~ \nonumber  \\
%\end{gather}
Note that if one is interested solely in the truncation of a given orthogonality center and keeping it at the same site, an even more efficient approach would be to perform an SVD on the matrix $C_{(q_k),(s_i Q_i \setminus q_k)} $. Again, the orthogonality of the states $\ket{q_k}$ and the orthogonality of the $U$ and $V^\dag$ matrices guarantee 
that the left and right vectors also form an orthogonal basis and hence Eq.~\ref{eq:schmidtTTN} is a true Schmidt 
decomposition. 
Note that this Schmidt decomposition separates all sites in segment $S^i_{q_k}$ as well as site $i$ to the 
rest of the lattice. From there it is straightforward to calculate the reduced density matrix for one of these two 
subsystems and approximate states by keeping only the largest eigenvalues in the spirit of DMRG.

\section{TDVP equations for Tree Tensor Networks}\label{sec:TDVP_TTN}
\begin{figure}
 \centering
  \parbox{8cm}{
\includegraphics{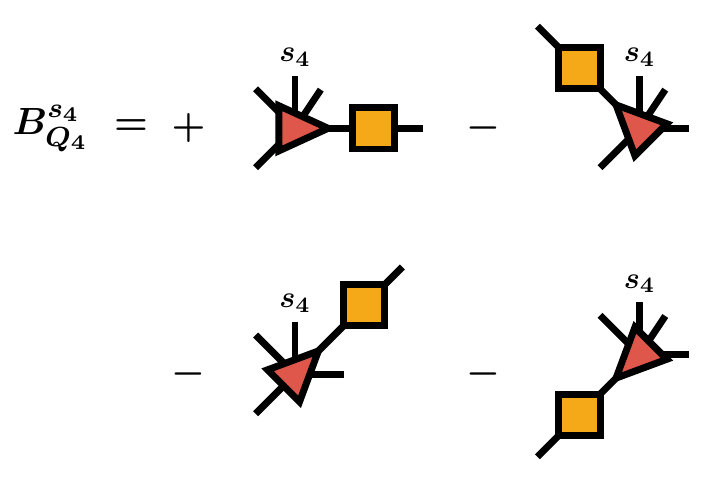} }
\qquad
\begin{minipage}{3cm}
\includegraphics[scale=0.7]{BaseTTN.pdf}
\end{minipage}

\caption{Definition of the vertical subspace for site $4$ in Eq.~\ref{eq:vertSubspace} with end point 
site $7$. The space of all tensors defining the kernel of the map from the tangent space to the physical hilbert space, 
is spanned by a matrix $X_{q_l' q_l}$ (yellow squares) for each link of the TTN. The physical index 
$s_4$ is labeled separately to distinguish it from the bond indices. The inset is a reminder of the TTN used in this section. }
\label{fig:VertSubspace}
\end{figure}
In this section, we generalize the derivation of the tangent space projector presented for MPS in 
Ref.~\cite{HaegemanTDVP1} to general TTNs. While the overall approach is very similar to the derivation for MPS, 
the lack of a clear start and end point in the tensor network geometry will make the derivation and the subsequent 
integration of the equations quite different from standard MPS. 
\\
TDVP amounts to the solution of the Schr{\"o}dinger equation in the space spanned by the tensor network 
without ever leaving this manifold (at least in its single-site variant). In TDVP, one solves a modified 
Schr{\"o}dinger equation by projecting its right-hand side onto the so-called tangent space:
\begin{equation}
 \label{eq:TDVP_schroedinger}
 \frac{d \ket{\psi [T] }}{dt} = -i \mathcal{P}_{T_{\ket{\psi[T]}}} H \ket{\psi[T]}.
\end{equation}
In the following, we want to find a representation of the tangent space projection operator 
$\mathcal{P}_{T_{\ket{\psi[T]}}}$, which not only depends on the current state $\ket{\psi[T]}$ but importantly also on 
the structure of the TTN. 
\subsection{Tangent Space Projector}
Any element of the tangent space $\ket{\Theta[B]}$ is parametrized by a set of tensors $B^{s_i}_{Q_i}$:
\begin{equation}
 \label{eq:def_tangentVector}
 \ket{\Theta[B]} =  \sum_{i=1}^N B^{s_i}_{Q_i} \frac{d \ket{\psi [T] }}{d T^{s_i}_{Q_i}}.
\end{equation}
Importantly, for each summand  we use the representation of the state $\ket{\psi[T]}$ in which site $i$ is the 
orthogonality center, i.e., all tensors $T^{s_j}_{Q_j}$ are orthogonalized towards site $i$ such that
\begin{equation}
 \ket{\Theta[B]} = \sum_{i=1}^N \sum_{ \stackrel{Q_i}{s_i} } B^{s_i}_{Q_i}  \ket{s_i} \ket{q^{[i]}_1 \cdots 
q^{[i]}_{r_i}}  \label{eq:tangentVector_qbasis}
\end{equation}
The gauge degree of freedom in the TTN, reflects itself in the tangent space that not all linearly independent choices 
of $B^{s_i}_{Q_i}$ result in different tangent vectors. Ref.~\cite{HaegemanTDVP1} solves this problem by first 
defining the so-called vertical subspace, i.e., all tensors $B^{s_i}_{Q_i}$ that give the zero-state $\ket{\Theta[B]} = 
0$ and hence define the kernel of the map from the tensors to the physical Hilbert space. Then, imposing a gauge 
prescription, they fix this kernel to a single 
element which guarantees that the resulting parametrization is unique. \\
In order to arrive at a result that resembles the MPS algorithm, we first need to define a fixed end point of the TTN with the 
restriction that it should be a leave. Note however that any site of the tensor network can be used as end point. Without loss of generality, we choose site $N$ as end point. The 
vertical subspace, i.e., all tensors $B^{s_i}_{Q_i}$ for which $\ket{\Theta[B]} =0$ can then be parametrized by 
matrices $X_{q_k' q_k}$ such that:
\begin{align}
 B^{s_i}_{Q_i} = \sum_{l=1}^{r_i} \sum_{q_l'} \left( T^{\mathcal{N}[q_l]} \right)^{s_i}_{q_1 \cdots q_l' \cdots
q_{r_i} } X_{q_l' q_l} \cdot \text{sgn}(q_l \to N) \nonumber \\%\left( -1 \right)^{\Theta( q_k \to N) }   \nonumber \\
\text{sgn}(q_l \to N) = \begin{cases} \mbox{1,} & \mbox{if } q_l \text{ points towards $N$} \\ 
\mbox{-1,} & \mbox{otherwise} \end{cases}. \label{eq:vertSubspace}
\end{align}
$(T^{\mathcal{N}[q_l]})^{s_i}_{Q_i}$ is the unique tensor of the state $\ket{\psi[T]}$ with site $i$ orthogonalized 
towards the neighbor on the other end of the link $q_l$. This definition of the vertical subspace is depicted in 
Fig.~\ref{fig:VertSubspace} for the tensor $B^{s_4}_{Q_4}$.
\\
The factor $\text{sgn}(q_l \to N) $ is $1$ if link $q_l$ points towards the end point and $-1$ otherwise. This 
construction guarantees that for any choice of $B^{s_i}_{Q_i}$ in the vertical subspace, 
$\ket{\Theta[B]} =0$, because the single term with positive sign $(q_l \to N)$ is exactly 
canceled by one negative term of its neighbor (since there $(q_l \nrightarrow N)$). Note that this definition of the 
vertical subspace reduces in the case of MPS to the definition used in Ref.~\cite{HaegemanTDVP1} if the 
right-most site of the MPS is chosen as the end point.
\\
To uniquely specify the kernel, we impose the following matrix-valued (with indices $q_k$ and $q_k'$) gauge fixing 
condition for the $B$-tensors of the tangent space:
\begin{align}
 \sum_{\stackrel{Q_i \setminus q_k }{s_i}} \bar{B}^{s_i}_{q_1 \cdots q_k' \cdots q_{r_i}} \cdot \left( T^{ 
\mathcal{N}[q_k]} \right)^{s_i}_{q_1 \cdots q_k \cdots q_{r_i}} = 0 \quad \forall \quad i  \neq N. 
\label{eq:tdvp_gaugefixing}
\end{align}
Again, the bar denotes complex conjugation. Above, $q_k$ is the single index pointing towards the end point $N$. These 
are $N-1$ matrix-valued constraints, for the $X$-matrices living on $L=N-1$ indices. This implies that no ambiguity is 
left in the definition of the kernel, if we choose $B$-tensors according to Eq.~\ref{eq:tdvp_gaugefixing}. 
\\
It also guarantees that the overlap between two tangent vectors reduces to a contraction over local 
tensors only:
\begin{equation}
 \langle \Theta[B'] | \Theta[B] \rangle = \sum_{i=1}^N \sum_{\stackrel{Q_i}{s_i}} \bar{B}^{s_i}_{Q_i} \cdot 
B^{s_i}_{Q_i}.
\end{equation}
Similar to MPS, we can now reformulate the projection problem of an arbitrary state $\ket{\Xi}$ onto the 
tangent space $\ket{\Theta[B]} = \mathcal{P}_{T_{\ket{\psi[T]}}} \ket{\Xi}$ as a minimization problem: 
\begin{equation}
 \min_{B} ||\ket{\Theta[B]} -\ket{\Xi} ||^2,
\end{equation}
under the constraints given by Eq.~\ref{eq:tdvp_gaugefixing}. With Eq.~\ref{eq:tangentVector_qbasis} and using a 
Lagrange multipliers $\lambda^{[i]}_{q_k q_k'}$ to account for the constraints, the minimization can be reformulated as:
\begin{align}
 \min_{B} \Bigg[ \sum_{i=1}^N  \sum_{\stackrel{Q_i}{s_i}} \left (  \bar{B}^{s_i}_{Q_i} \cdot 
B^{s_i}_{Q_i} - \bar{B}^{s_i}_{Q_i} \cdot F^{s_i}_{Q_i} - \bar{F}^{s_i}_{Q_i} \cdot B^{s_i}_{Q_i} \right ) \nonumber \\
- \sum_{i=1}^{N-1} \sum_{ q_k q_k'} \lambda^{[i]}_{q_k q_k'} \sum_{ \stackrel{Q_i 
\setminus q_k}{s_i} } \bar{B}^{s_i}_{q_1 \cdots q_k' \cdots q_{r_i}} \cdot \left( 
T^{\mathcal{N}[q_k]} \right)^{s_i}_{q_1 \cdots q_k \cdots q_{r_i}} \Bigg]
\end{align}
with $F^{s_i}_{Q_i} = \langle s_i q_1 \cdots q_{r_i} \ket{\Xi}$. The solution to this minimization problem can be found 
by setting the derivative with respect to $\bar{B}^{s_i}_{Q_i}$ as well as $\lambda^{[i]}_{q_k q_k'}$ to zero. Using 
some algebra we find the minimum for all sites $i \neq N$:
\begin{align}
  B^{s_i}_{Q_i} = &F^{s_i}_{Q_i} -  \sum_{ \stackrel{Q_i'' \setminus q_k'' , q_k'}{t}} \left( 
T^{\mathcal{N}[q_k]} 
\right)^{s_i}_{q_1 \cdots q_k' \cdots q_{r_i} } \nonumber \\
 & \cdot \left( T^{\mathcal{N}[q_k]} \right)^{t}_{q_1'' \cdots q_k' \cdots q_{r_i}'' } \cdot 
F^{t}_{q_1'' \cdots q_k \cdots q_{r_i}'' },
\end{align}
while for $i=N$ it is just $B^{s_N}_{Q_N} = F^{s_N}_{Q_N}$.
This  allows us to obtain a representation of the tangent space projector $\ket{\Theta[B]} = 
\mathcal{P}_{T_{\ket{\psi[T]}}} \ket{\Xi}$ as:
\begin{align}
 \mathcal{P}_{T_{\ket{\psi[T]}}} &= \sum_{i=1}^{N} \mathbb{1}_{s_i} \otimes \sum_{Q_i} \ket{q^{[i]}_1 \cdots 
q^{[i]}_{r_i}} \bra{q^{[i]}_1 \cdots q^{[i]}_{r_i}} \nonumber \\
&- \sum_{ <i,j>_{q_k} } \sum_{q_k q_k'} \ket{ q^{[j]\prime}_k } \bra{q^{[j]\prime}_k}  \otimes 
\ket{q^{[i]}_k} 
\bra{q^{[i]}_k}, \label{eq:TangentProjector}%\nonumber \\
%\ket{L_{q_k'}} &= \sum_{ \stackrel{Q_i \setminus q_k }{s_i} } \left( 
%T^{\mathcal{N}[q_k]} \right)^{s_i}_{q_1 \cdots q_k' \cdots q_{r_i}} \ket{s_i} \ket{q^{[i]}_1 \cdots 
%q^{[i]}_{k-1} q^{[i]}_{k+1} \cdots q^{[i]}_{r_i}}.
\end{align}
\begin{figure}
 \centering
\includegraphics{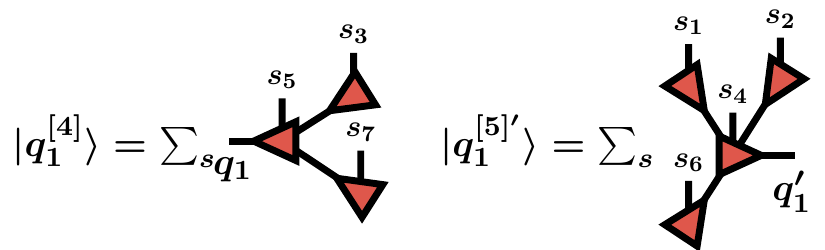}
\caption{Definition of the states used in the projection operator onto the link $q_k$ defined in the second line of 
Eq.~\ref{eq:TangentProjector} for link $q_1$ connecting sites $i=4$ and $j=5$. }
\label{fig:projBond}
\end{figure}

Where $ \sum_{ <i,j>_{q_k} }$ denotes a sum over all nearest neighbors $i$ and $j$ with the corresponding 
index $q_k$ connecting these two sites. The graphical representation of the states in the second line of the tangent 
space projector for the bond connecting sites $i=4$ and $j=5$ is shown in Fig.~\ref{fig:projBond}.
Formally, this result resembles the projection operator obtained for MPS~\cite{HaegemanTDVP1}. The 
first term with positive sign corresponds to the forward time propagation of the site tensor. The second term 
on the other hand is the evolution backwards in time of the bonds between two site tensors and is a direct 
consequence of the gauge fixing of the tangent vectors used in Eq.~\ref{eq:tdvp_gaugefixing}.

\subsection{Single-Site TDVP}
\begin{figure*}
 \centering
\includegraphics{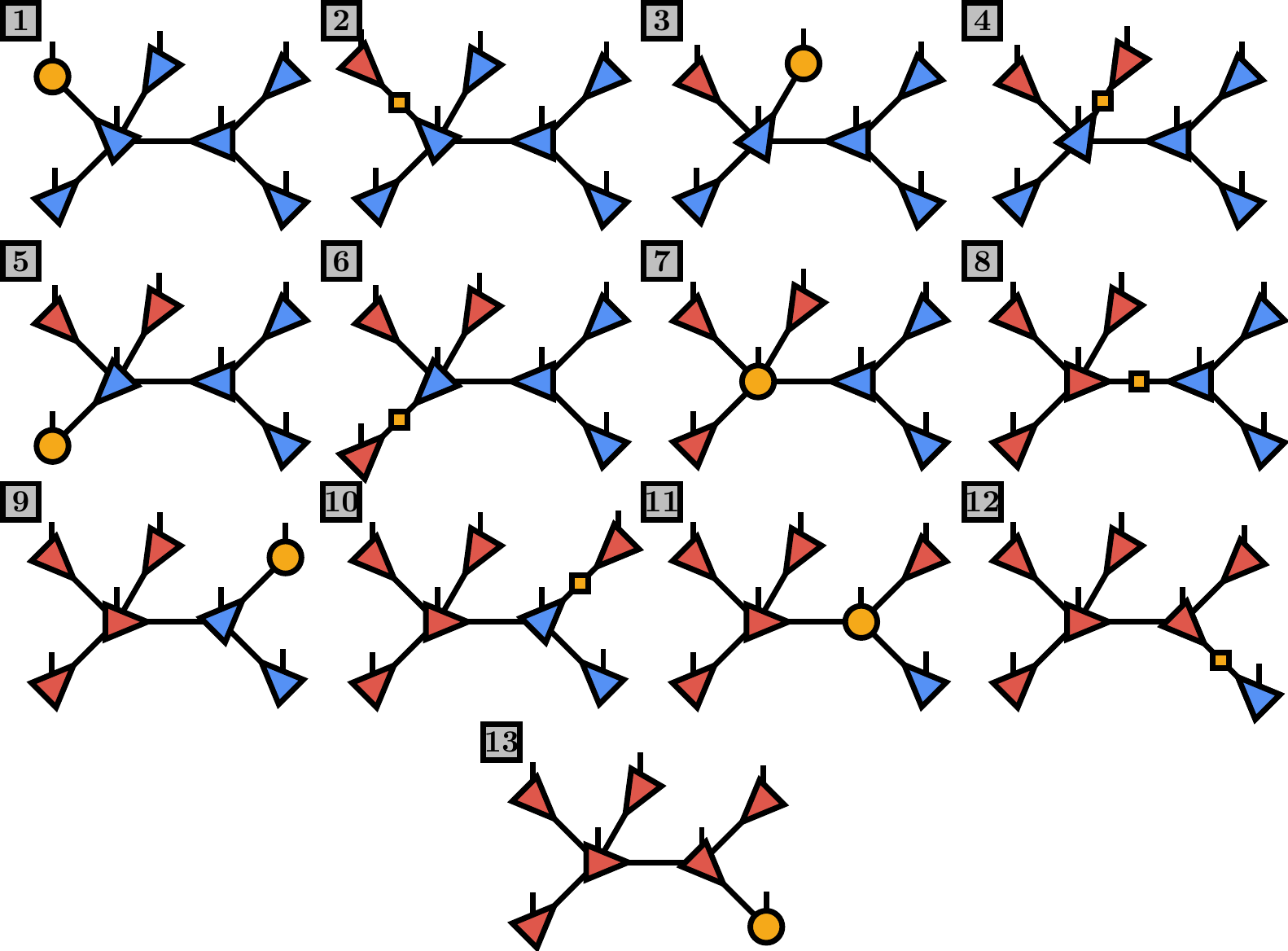}
\caption{Update sequence to perform a first-order single-site TDVP time step from time $t$ to $t+\dt$ for the TTN 
shown in Fig.~\ref{fig:TTN}. Start point is site $1$ and end point site $7$. Yellow denotes tensor that are updated 
in the current step. Red and blue tensors indicate whether this tensor is taken at time $t+\dt$ (red) or $t$ 
(blue). 
Triangles indicate the orthogonalization of each tensor. Updates on site-tensors are in forward direction (negative 
sign), while updates on bond-tensors are backwards time evolutions (positive sign in Eq~\ref{eq:localSchroedingerEq}) 
For a second order update, first perform all steps $(1)\to (13)$ with time step $\frac{dt}{2}$ in the order shown and 
then reapply them in the reverse order  $(13)\to (1)$, again with time step $\frac{dt}{2}$.   }
\label{fig:SingleSiteTDVP}
\end{figure*}

With the representation of the projection operator in Eq.~\ref{eq:TangentProjector}, we can go back to the projected 
time dependent Schr{\"o}dinger 
equation (Eq.~\ref{eq:TDVP_schroedinger}) and integrate each term one by one using Trotter 
breakups~\cite{SuzukiDecomp}. First, let us discuss a first-order update, which can later easily be modified to perform a second order integration. Since each term in the projection operator keeps all but one tensor fixed, the integration can be performed locally.  Therefore, we define effective Hamiltonians for the sites $i$ and 
for the links $q_k$:
\begin{subequations}
\begin{eqnarray}
 H_{ (s_i Q_i), (s_i' Q_i') } = \langle s_i q^{[i]}_1 \cdots q^{[i]}_{r_i} | H| s_i' q^{[i]\prime}_1 \cdots q^{[i]\prime}_{r_i} \rangle \label{eq:onesiteHeff} \\
 K_{(q^{[i]}_k q^{[j]}_k) (q^{[i]\prime}_k q^{[j]\prime}_k )} = \langle q^{[i]}_k q^{[j]}_k | H| q^{[i]\prime}_k q^{[j]\prime}_k \rangle \label{eq:zerositeHeff} 
\end{eqnarray}
\end{subequations}
and solve equations of the form:
\begin{align}
 \mathbf{ \dot{A} } &= \pm i H^{\text{eff}} \cdot \mathbf{ A } \nonumber \\
 \mathbf{A}(t+\dt) &= e^{ \pm i H^{\text{eff}} \dt} \mathbf{A}(t), \label{eq:localSchroedingerEq}
\end{align}
where $ \mathbf{A}$ is either a site-tensor or a link tensor and $H^{\text{eff}}$ either $H_{ (s_i Q_i), 
(s_i' Q_i') } $ (negative sign) or $K_{(q^{[i]}_k q^{[j]}_k) (q^{[i]\prime}_k q^{[j]\prime}_k )}$ (positive sign). In 
matrix form, the exponential of these effective Hamiltonians can be efficiently calculated using Krylov exponentiation.
\\
A full TDVP step is then given by a series of $N-1$ local updates of a site tensor and the corresponding link tensor 
connecting the site to the end point as shown below. The single local update on site $i$ and link $q_k$ is
%% single update on site i 
\begin{itemize}
 \item Orthogonalize the TTN such that site $i$ is the orthogonality center.
 \item Calculate the one-site effective Hamiltonian $H^{\text{eff}} = H_{ (s_i Q_i), (s_i' Q_i') }$ 
(Eq.~\ref{eq:onesiteHeff}) 
and forward time evolve (negative sign) according to Eq.~\ref{eq:localSchroedingerEq} with $\mathbf{A} = 
C^{s_i}_{Q_i}$. 
If site $i$ is the chosen end point, stop here; otherwise continue.
 \item Reshape the time evolved tensor into a matrix $C^{s_i}_{Q_i} = C_{ (s_i Q_i\setminus q_k), (q_k) 
}$ and perform an SVD (QR-decomposition suffices) $C_{ (s_i Q_i\setminus q_k), (q_k) } = 
\sum_{q^{[i]}_k} U_{ (s_i Q_i\setminus q_k), (q^{[i]}_k) } \cdot \underbrace{S_{q^{[i]}_k} \cdot (V^\dag)_{ 
(q^{[i]}_k), (q_k)}  }_{L_{q^{[i]}_k q_k}}$. As usual, take the $U$-tensor as new tensor on site $i$. 

 \item Calculate the link effective Hamiltonian $H^{\text{eff}} =K_{(q^{[i]}_k q^{[j]}_k) (q^{[i]\prime}_k 
q^{[j]\prime}_k )}$ (Eq.~\ref{eq:zerositeHeff}) for link $q_k$. To do so, use the 
time evolved tensor obtained in the previous step for site $i$. Then evolve tensor $\mathbf{A} = L_{q^{[i]}_k 
q_k} \equiv L_{q^{[i]}_k q^{[j]}_k}$ from the previous step backwards in time (positive sign) according to 
Eq.~\ref{eq:localSchroedingerEq}. Finally, absorb the $C$-tensor onto the neighbor of site $i$ along $q_k$ by 
multiplying it onto its site tensor.
\end{itemize}
%% full one-site TDVP 
A full TDVP time step can then achieved by the following sweeping procedure:
\begin{enumerate}
 \item Choose a start and an end point; initialize site $i$ as the chosen start point.
 \item Perform the following steps until $i$ is the chosen end point:
 \begin{itemize}
 \item Find the link $q_k \in Q_i$ that connects site $i$ to the end point. 
 \item If any tensor attached to the other links $Q_i \setminus q_k$ has not been updated, choose one of these 
links and choose one of the leaves attached to the corresponding segment of the TTN as new site $i$.
 \item Otherwise, perform a local update on site $i$ as described above and choose the neighbor of site $i$ along link 
$q_k$ as new site $i$.
 \end{itemize}
 \item Perform one last local update for the endpoint $i=N$ as described above.
\end{enumerate}
A depiction of the sweeping order for the TTN in Fig.~\ref{fig:TTN} is shown in Fig.~\ref{fig:SingleSiteTDVP}.
The procedure described above defines a first-order time step. A second-order method can easily be obtained by 
performing the first order time step with $\frac{\dt}{2}$ and then simply perform the exact same steps in reverse order 
corresponding to repeated second order Trotter breakups $e^{\tau(A+B+C)} = 
e^{\frac{\tau}{2}C} e^{\frac{\tau}{2}B}e^{\tau A} e^{\frac{\tau}{2}B} e^{\frac{\tau}{2}C}$ used on 
Eq.~\ref{eq:TangentProjector}. Importantly, this means that during the local update, the link update has to be performed 
\emph{before} the site update (see also caption of Fig.~\ref{fig:SingleSiteTDVP}). 
\\
Note that for a given TTN, there can be several versions of this algorithm depending on the sequence of chosen indices 
in step 2. Very often though, the TTN structure itself defines some natural order when to time evolve which sites, as 
we will see in the next section for the FTPS tensor network.

\subsection{Two-Site TDVP}
\begin{figure*}
 \centering
\includegraphics{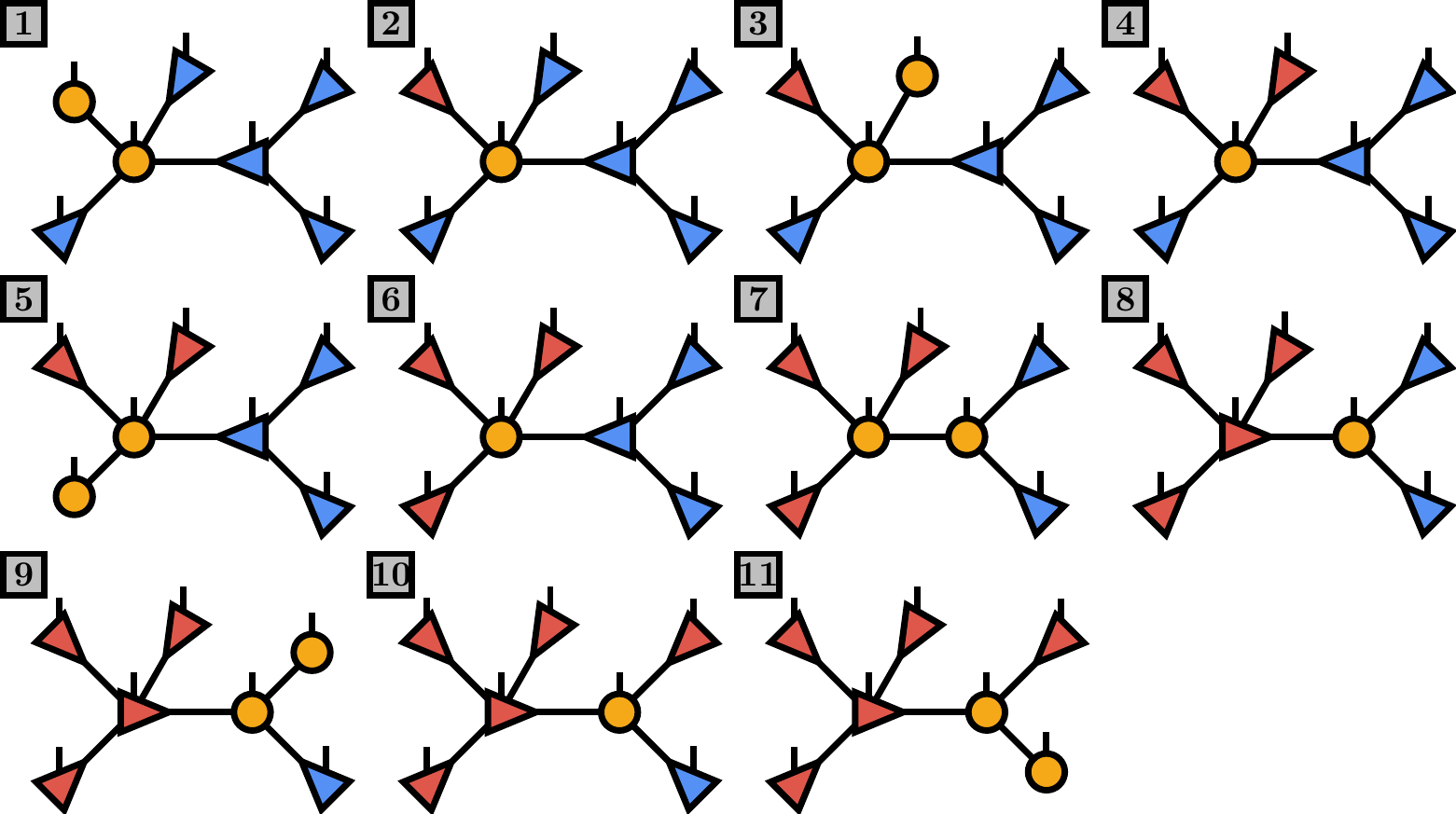}
\caption{Update sequence to perform a first-order two-site TDVP time step from time $t$ to $t+\dt$ for the TTN 
shown in Fig.~\ref{fig:TTN}. Start point is site $1$ and end point site $7$. Yellow denotes tensors that are updated 
in the current step. Red and blue tensors indicate whether this tensor is taken at time $t+\dt$ (red) or $t$ 
(blue). 
Triangles indicate the orthogonalization of each tensor. Updates on two-site tensors are in forward direction (negative 
sign), while updates on a single site are backwards time evolutions (positive sign in Eq.~\ref{eq:localSchroedingerEq}) 
For a second-order update, first perform all steps $(1)\to (11)$ with time step $\frac{dt}{2}$ in the order shown and 
then reapply them in the reverse order  $(11)\to (1)$, again with time step $\frac{dt}{2}$.   }
\label{fig:TwoSiteTDVP}
\end{figure*}

It is also straightforward to generalize the single-site TDVP approach presented above to a two-site 
TDVP integration scheme which allows to dynamically adapt the necessary bond dimensions. To do so, we need to define 
the two-site effective Hamiltonian $H^{\text{2-site}}$ for two sites $i$ and $j$ connected by the index $q_l$:
\begin{align}
  H^{\text{2-site}} &= \langle s_i s_j Q^{\text{red}}_i Q^{\text{red}}_j |H| s_i' s_j' Q^{\text{red}}_i 
Q^{\text{red}}_j  \rangle \nonumber \\
  \ket{ Q^{\text{red}}_i } &= \bigotimes_{q_n \in Q_i \setminus q_l} \ket{q^{[i]}_n} \nonumber \\
  \ket{ Q^{\text{red}}_j } &= \bigotimes_{q_n \in Q_j \setminus q_l} \ket{q^{[j]}_n}. \label{eq:heff2site}
\end{align}
With this, only small modifications to the algorithm presented above are necessary. The single update for 
sites $i$ and $j$ sharing link $q_l$ becomes:

%% single update on site i and j
\begin{itemize}
 \item Orthogonalize the TTN such that site $i$ is the orthogonality center. 
 \item Calculate the two-site effective Hamiltonian $H^{\text{2-site}}$ according to Eq.~\ref{eq:heff2site}. 
and forward time evolve (negative sign) with $\mathbf{A} = \sum_{q_l} C^{s_i}_{Q_i} T^{s_j}_{Q_j}$.
 \item Reshape the time evolved tensor into a matrix $A_{ (s_i Q_i\setminus q_l), (s_j Q_j\setminus 
q_l) }$ and perform an SVD $A_{ (s_i Q_i\setminus q_l), (s_j Q_j\setminus 
q_l) } = \sum_{q_l} U_{ (s_i Q_i\setminus q_l), (q_l) } \cdot \underbrace{S_{q_l} 
\cdot (V^\dag)_{ (q_l), (s_j Q_j\setminus q_l)}  }_{C^{s_j}_{Q_j}}$. In this step one can also truncate the smallest Schmidt values. As usual, keep the $U$-tensor to 
update site $i$ and $C^{s_j}_{Q_j}$ as site tensor on site $j$, shifting the orthogonality center to site $j$. If site $j$ is the chosen end point, stop here; 
otherwise continue.
 \item Calculate the one-site effective Hamiltonian $H^{\text{eff}} = H_{ (s_j Q_j), (s_j' Q_j') }$ 
(Eq.~\ref{eq:onesiteHeff}) 
for site $j$. To do so, use the time-evolved tensor obtained in the previous step for site $i$. Then evolve tensor 
$\mathbf{A} = C^{s_j}_{Q_j}$ backwards in time (positive sign in Eq.~\ref{eq:localSchroedingerEq}).
\end{itemize}
A full two-site TDVP step can then be performed by:
\begin{enumerate}
 \item Choose a start and end point. Initialize site $i$ as the chosen start point.
 \item Perform the following steps until $i$ is the chosen end point:
 \begin{itemize}
    \item Find the link $q_k$ and the corresponding neighbor $j$ that connects site $i$ to the end point. 
    \item If any tensor attached to the other links $Q_i \setminus q_k$ has not been updated, choose one of these 
links and choose one of the leaves attached to the corresponding segment of the TTN as new site $i$.
    \item Otherwise, perform a local update on site $i$ and $j$ as described above and go to site $j$, i.e., $i \to 
j$.
  \end{itemize}
\end{enumerate}
A depiction of the necessary sweeping order of this two-site scheme is shown in Fig.~\ref{fig:TwoSiteTDVP}

\section{TDVP for FTPS}\label{sec:TDVP_FTPS}
\begin{figure*}
    \centering
    \includegraphics{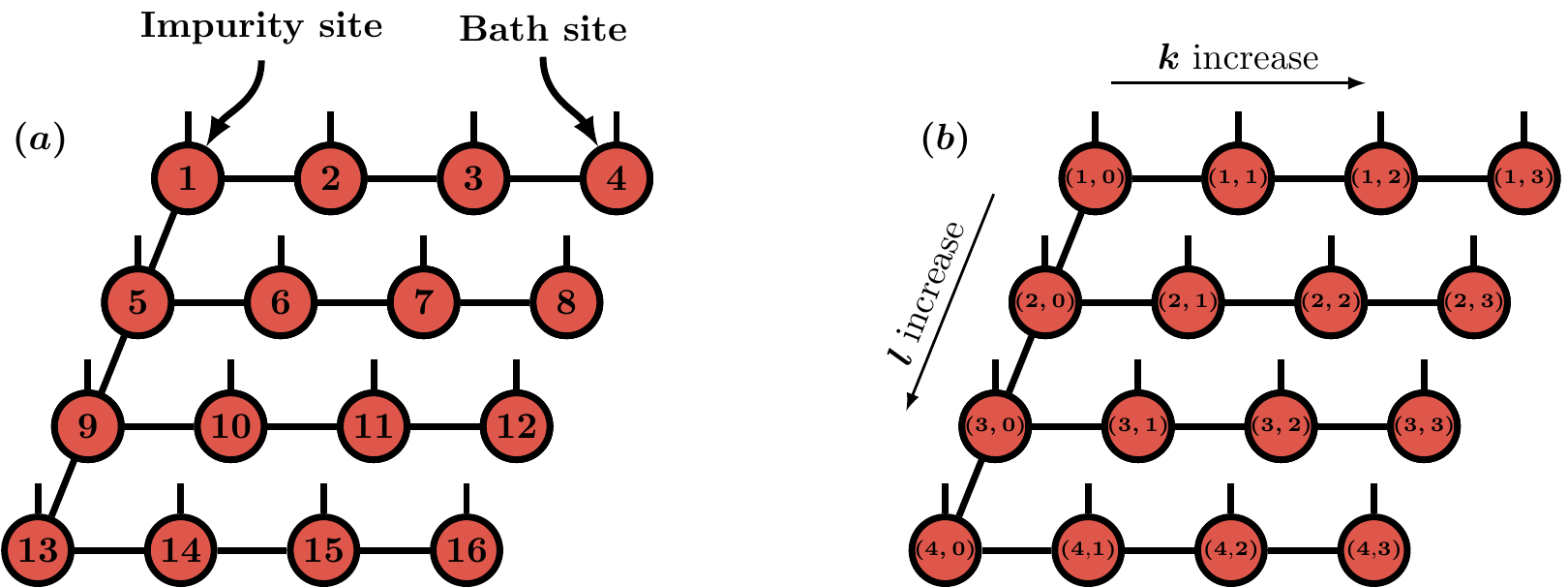}
    \caption{Graphical representation of a FTPS tensor network for a two orbital model. For each orbital, we use two 
    chains, one for each spin-species. \emph{(a)} one way to label the sites is just to numerate them in ascending 
    order. 
    \emph{(b)} a different way to label sites is to specify the chain (orbital $m$ and spin $\sigma$) as well as an 
    index (bath index $k$). This way to label sites resembles the labels used for the operators of the Hamiltonian in 
    Eq.~\ref{eq:H_AIM}.} 
    \label{fig:FTPS}
\end{figure*}

An FTPS is a special TTN designed to efficiently encode states of multi-orbital Anderson 
Impurity Models (AIMs). An AIM consists of an interacting impurity coupled to a bath of free fermions with Hamiltonian
\begin{align}
    H &= \Hloc + \Hbath + \Hhyb \nonumber \\
    \Hloc~~ &=   \sum_{m \sigma} \epsilon_{m \sigma 0} n_{m \sigma 0} + H_{\text{int}}. \label{eq:H_AIM} \nonumber \\
    \Hbath &=  \sum_{m\sigma} \sum_k \epsilon_{m\sigma k} n_{m\sigma k} \nonumber \\
    \Hhyb~  &= \sum_{m\sigma} \sum_k V^{[k]}_{m\sigma} \left( c_{m\sigma 0}^\dag 
    c_{m\sigma k}+ \text{h.c.} \right)
\end{align}
$c_{m\sigma k}^\dag$ ($c_{m \sigma k}$) creates (annihilates) an electron in chain $m$ with spin $\sigma$ on 
site $k$, where $k=0$ denotes the impurity site (see Fig.~\ref{fig:FTPS} (b)). $n_{m\sigma k}$ are the corresponding 
particle number operators. 
$H_{\text{int}}$ is the interaction Hamiltonian that only couples impurity degrees of freedom and for which we choose 
the Kanamori Interaction~\cite{Kanamori,FTPS} without the spin-flip and pair-hopping terms parametrized by two interaction strengths $U$ and $J$
\begin{equation}
H_{\text{int}} = U \sum_m n_{m \uparrow 0} n_{m \downarrow 0} + (U-2J) \sum_{m'>m \sigma} n_{m \sigma 0} n_{m' \bar{\sigma} 0} 
+ (U-3J) \sum_{m'>m \sigma} n_{m \sigma 0} n_{m' \sigma 0} 
\end{equation}
where $\bar{\sigma}$ is the opposite spin direction of $\sigma$.
In the following, we will use a combined index $l = (m \sigma)$ to denote the orbital and spin-degrees of freedom.
\\
For a single orbital, an FTPS reduces to a MPS, while for multiple orbitals it has tensors with 
three link indices as depicted in Fig.~\ref{fig:FTPS} for a two-orbital model. It consists of a single MPS-like chain 
for the bath tensors of each orbital/spin and impurity tensors connecting the different chains. An FTPS for a 
$N_{\text{orb}}$-orbital AIM has a total of $N_{C} = 2N_{\text{orb}}$ chains. For simplicity we assume that each chain 
has the same number of bath sites $N_b$. \\~\\
According to the algorithm presented in the previous section, we first need to choose a start and end point. We choose 
to start at the outermost bath site of the first chain (site $4$ in Fig.~\ref{fig:FTPS} (a)) and the outermost bath 
site of the last chain as end point (site $16$ in Fig.~\ref{fig:FTPS} (a)). To actually perform the time evolution, we 
choose to employ a hybrid TDVP scheme using 2-site TDVP for the bath tensors as well as for the bath-impurity link, and 
1-site TDVP for the impurity tensors itself and the corresponding impurity-impurity links. We choose to use 1-site TDVP 
for the impurity links, since 2-site TDVP becomes computationally expensive, since one would have to deal with tensors 
with four link indices. This leads to the following algorithm for a single time step: 
\begin{enumerate}
    \item For $l = 1 : N_{C}-1$ perform the following steps:
    \begin{itemize}
    \item For $k= N_b:1$:
      \begin{itemize}
      \item Perform a two-site step on sites $i=(l,k)$ and $j=(l,k-1)$ (see Fig.~\ref{fig:FTPS} for the definition of 
	the site-labeling).
      \end{itemize}
    
    \item Perform a one-site step on the impurity tensor $i=(l,0)$; $q_k$ connects site $(l+1,0)$
    \end{itemize}
    \item $l = N_{C}$, for $k = 0:N_b-1$ perform the following steps:
    \begin{itemize}
    \item Perform a two-site step on sites $i=(l,k)$ and $j=(l,k+1)$.
    \end{itemize}
\end{enumerate}
For the actual calculations, we apply the second order version of this algorithm by using only the half time step and 
reapplying each step in reverse order. Again, this also means that the order in the local updates changes. Note that the backwards propagation during the two-site involving an impurity site cancels with the subsequent forwards time evolution of the one-site step on the same impurity tensor. Therefore, these two steps can be omitted.

\begin{figure}
	\centering
	\input{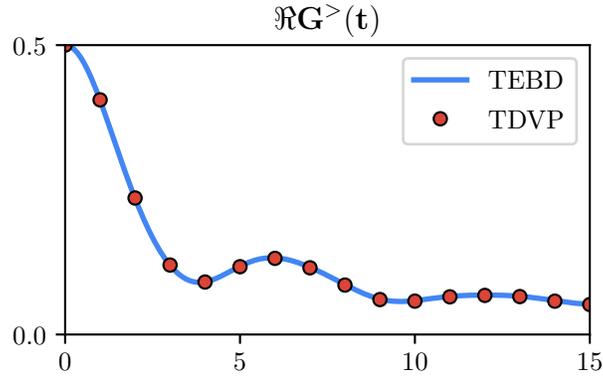}
	\caption{Comparison of the impurity greater Greens function $G^>(t)$ for a two orbital model between the TEBD 
time-evolution approach used in Ref.~\cite{FTPS} and TDVP. The calculation was performed for a spin- and 
orbital 
	degenerate model using $9$ bath sites per orbital and spin with parameters $\epsilon_k = -0.8, -0.6, \cdots, 
	0.6, 0.8$ and $V^{[k]}_l = 0.1 ~ \forall ~ k,l$. Therefore, only diagonal entries of the Green's function are 
    non-zero and for the diagonals there is only one independent function, i.e., $ G^{>}_{ l',l }(t) = 
    \delta_{l'l} G^{>}(t)$. Interaction parameters were $U=1$, $J=0.1$. The impurity on-site energy was chosen to obtain particle hole symmetry, i.e., 
    $\epsilon_{m\sigma0}=-\frac{3U-5J}{2}$. The time step for TEBD was $\dt = 0.01$ and for TDVP $\dt = 
    0.1$, since TDVP generally allows to use larger time steps~\cite{StarVSChainsPaper}. Truncated weight (sum of all truncated Schmidt values) for TEBD was $10^{-12}$ without restricting the bond dimension, and for  TDVP it was $10^{-9}$ for all links except the impurity-impurity links which were not truncated, but restricted to a maximal dimension of 50. }
	\label{fig:FTPS_TDVPvsSWAP}
\end{figure}

As a first demonstration of this algorithm, let us compare the TDVP time evolution to the TEBD-like 
approach used in Refs.~\cite{FTPS,THESIS}. Therefore, we look at the greater Greens function of the impurity 
defined by:
\begin{equation}
 G^{>}_{ l', l }(t) = \langle \psi_0 | c_{l' 0} e^{-iHt}  c_{l 0}^\dag  | \psi_0 \rangle 
e^{i E_0 t}.
\end{equation}
$\ket{\psi_0}$ is the ground state of Hamiltonian $H$ with ground state energy $E_0$. For 
a degenerate two orbital model, Fig.~\ref{fig:FTPS_TDVPvsSWAP} shows that the TDVP time evolution indeed produces the 
correct result. In a recent publication, the authors have shown, that for diagonal hybridizations, TDVP has larger 
errors than TEBD for the bath geometry chosen here~\cite{StarVSChainsPaper}. This means that for such systems, the TEBD 
approach is most likely preferable over TDVP. For more involved baths on the other hand, TEBD can become difficult 
to formulate as discussed next.
\\~\\
One of the major advantages of TDVP is that it allows to perform the time evolution for arbitrary couplings 
in the Hamiltonian between the sites, as long as an \emph{MPO} with the same tensor network structure as the state can 
be found. Eq.~\ref{eq:H_AIM} is in fact not the most general AIM, since the bath only couples \emph{diagonally} to its 
impurity. Often, one is also interested in so-called off-diagonal hybridizations which can be encoded as hoppings from 
impurity $l$ to a different bath $l'$. Therefore, we can account for off-diagonal hybridizations 
by replacing the hybridization terms in Eq.~\ref{eq:H_AIM} with:
\begin{equation}
 \sum_{ l l' k } V^{[k]}_{ l l' } (c_{l 0}^\dag c_{l' k} +\text{h.c}).
\end{equation}
It turns out that for each $k$, the matrix $V^{[k]}_{ l l' }$ can be chosen as a lower-triangular matrix. This means, 
that for a spin-symmetric, two-orbital model there are three free parameters for each value of 
$k$ (instead of two for the diagonal hybridization). 
\begin{figure*}
	%remove the two lines addlegendimage
	%
	\centering
	\input{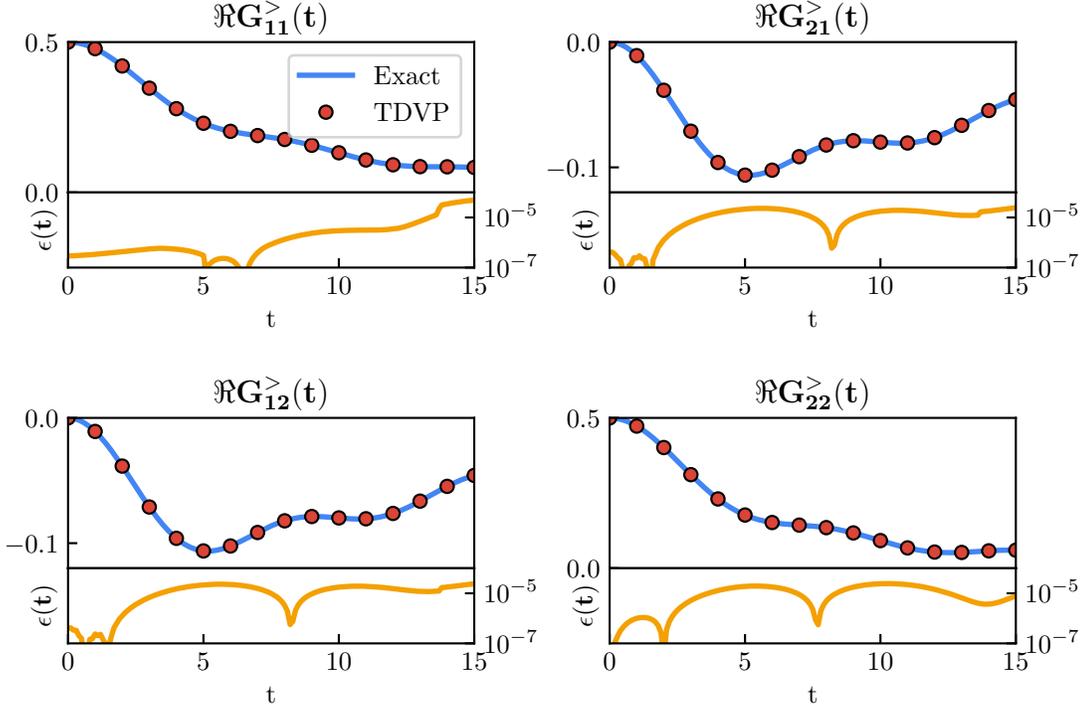}
	\caption{Comparison of impurity greater Greens function $G_{ll'}^>(t)$ for a two-orbital model between 
	TDVP and the exact solution. In each segment, the top panel shows the Green's function itself, while the bottom 
panel shows the absolute value of the difference $\epsilon(t) = | G^>_{\text{exact}}(t) - G^>_{\text{TDVP}}(t) |$, 
i.e., the numerical error. We obtained the exact solution from 
diagonalization of the hopping matrix at $U=J=0$. The calculation was performed for a spin-degenerate model using $9$ 
bath sites per orbital and spin. We allowed off-diagonal hopping terms only between the orbital degrees of freedom, 
i.e., $V^{[k]}_{(m\sigma)(m' \sigma')} = \delta_{\sigma \sigma'} V^{[k]}_{m m'}$ and therefore $G_{(m\sigma)(m' 
\sigma')} = \delta_{\sigma \sigma'} G_{m m'}$. The parameters were $\epsilon_k = -0.8, -0.6, \cdots, 	0.6, 0.8$ for 
all orbitals, diagonal hybridizations $V^{[k]}_{mm} = 0.1 ~ \forall ~ k,l$ and off-diagonal hybridizations $V^{[k]}_{m 
m'} = 0.05 ~ \forall ~ k$ for $m = 2, m' = 1$. These off-diagonal terms correspond to a hopping processes from the 
impurity of orbital $2$ to the bath of orbital $1$. The TDVP time step was chosen $\Delta t = 0.1$ and on-site energies 
were $\epsilon_{m\sigma0}=0$. Note that the off-diagonal hybridizations break the orbital degeneracy, albeit for 
the parameters chosen only slightly and the differences between the two orbitals are barely visible. Truncated weight (sum of all truncated Schmidt values) during DMRG and the time evolution was $10^{-9}$, except during the time evolution of the impurity impurity links where no truncation was performed. During DMRG as well as time evolution, the impurity-impurity links were restricted to 140. }
	\label{fig:FTPS_TDVPoffDiagvsExact}
\end{figure*}

\begin{figure}
\centering
\parbox{7cm}{
\input{Fig12.pgf}
\label{fig:2figsA}}
\qquad
\begin{minipage}{7cm}
\input{Fig13.pgf}
\label{fig:2figsB}
\end{minipage}
\caption{ Error as a function of control parameters for the same AIM used in Fig.~\ref{fig:FTPS_TDVPoffDiagvsExact}. We plot the maximum value of the error obtained in all four Green's functions $max(\epsilon(t))$.\\
 \emph{Left:} As function of step size $\Delta t$, the error shows the expected scaling $\sim (\Delta t)^2$ for larger values of $\Delta t$. The deviations for smaller values can be explained by the other sources of error, like the truncation during time evolution and a not perfect representation of the ground state. Additionally, we found TDVP in the star geometry to be quite sensitive to a too small time step in combination with a too large truncation. The parameters used in the truncation of the tensor network were exactly the same as discussed above.  \\\emph{Right:} Error as a function of impurity-impurity bond dimension. All other parameters were the same as above.}
\label{fig:FTPS_TDVPoffDiag_ErrorScaling}
\end{figure}
As a second demonstration of the TDVP approach for FTPS we calculate the $2\times 2$ matrix of the greater 
Green's function of such a spin symmetric two-orbital model. Since the TEBD approach we compared with in 
Fig.~\ref{fig:FTPS_TDVPvsSWAP} is difficult to generalize to such off-diagonal hybridizations, we 
perform the calculation in the non-interacting case $U=J=0$ and note that for tensor network based approaches this is a 
highly non-trivial situation. This is because the bipartitions defined by the links of the FTPS structure have non-trivial entanglement also for non-interacting systems, and the off-diagonal hoppings $V^{ [k]}_{l l'}$ for $l \neq l'$ introduce entanglement \emph{between} the orbitals, i.e., 
non-trivial links between the impurities. The results of such a comparison can be seen in Fig.~\ref{fig:FTPS_TDVPoffDiagvsExact}. Having access to the exact solution, we also plot the difference between the exact and numerical Green's functions in the bottom panels. Again we find very good agreement between TDVP and the reference calculations. 
\\
Finally let us demonstrate that the results indeed converge with respect to the control parameters. The left plot of Fig.~\ref{fig:FTPS_TDVPoffDiag_ErrorScaling} shows the scaling of the error as a function of $\Delta t$ and we indeed observe the expected $~\sim \dt^2$ behavior at larger values of $\dt$. The deviation of this behavior at smaller $\dt$ can be understood from the additional errors due to the truncations of the tensor network in the ground state as well as during time evolution. Additionally, we frequently observed that when TDVP is used in the star-geometry representation of the bath (with long-range couplings $V_{m \sigma}$), a good balance between truncation and $\dt$ is necessary. Surprisingly we found that it is often advantageous to use rather large time steps compared to what one would use in TEBD calculations. In the right plot of Fig.~\ref{fig:FTPS_TDVPoffDiag_ErrorScaling} we show the convergence of the error as a function of dimension of the impurity-impurity links which is usually the bottle-neck of FTPS calculations as these links need to transport the entanglement between the different orbitals $l$. Also here, we observe convergence with the control parameter, showing that TDVP indeed can be efficiently used to account for off-diagonal hybridizations.

\section{Conclusion}
We presented a generalization of the Time Dependent Variational Principle (TDVP) to general loop-free tensor 
networks (TTNs). The major advantage of TDVP over the commonly used TEBD approach is that the latter is often difficult 
to implement if long-range couplings are present in the Hamiltonian. TDVP on the other hand allows to perform the time 
evolution (either in imaginary- or real-time) for any Hamiltonian for which a representation in the same TTN structure 
can be found, which is often possible for long-range couplings. Using a similar derivation as in 
Ref.~\cite{HaegemanTDVP1}, we were able to find the projection operator onto the tangent space for any TTN - the 
central object in TDVP. Integrating the terms in the tangent space projector one after the other, equivalent to a 
Suzuki Trotter breakup, we were able to formulate TDVP in its single-site as well as two-site variant. We then applied 
TDVP to the FTPS tensor network which is a TTN especially suited for multi-orbital Anderson impurity models. For FTPS, 
TDVP is particularly appealing if the hybridizations with the bath are off-diagonal. In DMFT calculations, off-diagonal 
hybridizations are of significance to account for spin-orbit 
coupling effects as well as distortions of the crystal lattice. We verified the TDVP approach by comparing first to TEBD 
using a diagonal bath including interactions, and second to the exact solution in the non-interacting case for an 
off-diagonal bath.
\\
When finalizing this manuscript we became aware of an independent publication by Kohn et al.~\cite{Kohn_TTNTDVP}, 
describing the TDVP applied to a TTN for periodic boundary conditions in a one-dimensional system.
\\ ~ \\
The authors would like to thank Florian Maislinger, Hans Gerd Evertz and Jutho Haegeman  for fruitful discussions. This 
work was supported by the Austrian Science Fund (FWF) through the START program Y746, as well as by NAWI Graz.

\bibliography{TTN_biblio}% Produces the bibliography via BibTeX.

\begin{thebibliography}{10}
\providecommand{\url}[1]{\texttt{#1}}
\providecommand{\urlprefix}{URL }
\expandafter\ifx\csname urlstyle\endcsname\relax
  \providecommand{\doi}[1]{doi:\discretionary{}{}{}#1}\else
  \providecommand{\doi}{doi:\discretionary{}{}{}\begingroup
  \urlstyle{rm}\Url}\fi
\providecommand{\eprint}[2][]{\url{#2}}

\bibitem{WhiteDMRG}
S.~R. White,
\newblock \emph{{Density matrix formulation for quantum renormalization
  groups}},
\newblock Phys. Rev. Lett. \textbf{69}, 2863 (1992),
\newblock \doi{10.1103/PhysRevLett.69.2863}.

\bibitem{SchollwoeckDMRG_MPS}
U.~{Schollw{\"o}ck},
\newblock \emph{{The density-matrix renormalization group in the age of matrix
  product states}},
\newblock Ann. Phys. \textbf{326}, 96 (2011),
\newblock \doi{10.1016/j.aop.2010.09.012}.

\bibitem{Ostlund_DMRGprodMPS}
S.~\"Ostlund and S.~Rommer,
\newblock \emph{Thermodynamic limit of density matrix renormalization},
\newblock Phys. Rev. Lett. \textbf{75}, 3537 (1995),
\newblock \doi{10.1103/PhysRevLett.75.3537}.

\bibitem{Dukelsky_RelationDMRGMPS}
J.~Dukelsky, M.~A. Martin-Delgado, T.~Nishino and G.~Sierra,
\newblock \emph{Equivalence of the variational matrix product method and the
  density matrix renormalization group applied to spin chains},
\newblock EPL (Europhysics Letters) \textbf{43}(4), 457 (1998).

\bibitem{VerstraeteCirac_PEPS_2004}
F.~{Verstraete} and J.~I. {Cirac},
\newblock \emph{{Renormalization algorithms for Quantum-Many Body Systems in
  two and higher dimensions}},
\newblock arXiv e-prints cond-mat/0407066 (2004),
\newblock \eprint{cond-mat/0407066}.

\bibitem{MurgVerstraete_PEPS2_2007}
V.~Murg, F.~Verstraete and J.~I. Cirac,
\newblock \emph{Variational study of hard-core bosons in a two-dimensional
  optical lattice using projected entangled pair states},
\newblock Phys. Rev. A \textbf{75}, 033605 (2007),
\newblock \doi{10.1103/PhysRevA.75.033605}.

\bibitem{VidalMera}
G.~Vidal,
\newblock \emph{Entanglement renormalization},
\newblock Phys. Rev. Lett. \textbf{99}, 220405 (2007),
\newblock \doi{10.1103/PhysRevLett.99.220405}.

\bibitem{Delgado_DMRGDendrimers_2002}
M.~A. Mart\'{\i}n-Delgado, J.~Rodriguez-Laguna and G.~Sierra,
\newblock \emph{Density-matrix renormalization-group study of excitons in
  dendrimers},
\newblock Phys. Rev. B \textbf{65}, 155116 (2002),
\newblock \doi{10.1103/PhysRevB.65.155116}.

\bibitem{Depenbrock_BetheLatticeImagTEBD_2013}
S.~Depenbrock and F.~Pollmann,
\newblock \emph{Phase diagram of the isotropic spin-$\frac{3}{2}$ model on the
  $z=3$ bethe lattice},
\newblock Phys. Rev. B \textbf{88}, 035138 (2013),
\newblock \doi{10.1103/PhysRevB.88.035138}.

\bibitem{Otsuka_DMRGTTN_1996}
H.~Otsuka,
\newblock \emph{Density-matrix renormalization-group study of the
  spin-$\frac{1}{2}$ $\mathrm{XXZ}$ antiferromagnet on the bethe lattice},
\newblock Phys. Rev. B \textbf{53}, 14004 (1996),
\newblock \doi{10.1103/PhysRevB.53.14004}.

\bibitem{Friedman_DMRGCayleyTree_1997}
B.~Friedman,
\newblock \emph{A density matrix renormalization group approach to interacting
  quantum systems on cayley trees},
\newblock Journal of Physics: Condensed Matter \textbf{9}(42), 9021 (1997),
\newblock \doi{10.1088/0953-8984/9/42/016}.

\bibitem{Gerster_UnconstrainedTree_2014}
M.~Gerster, P.~Silvi, M.~Rizzi, R.~Fazio, T.~Calarco and S.~Montangero,
\newblock \emph{Unconstrained tree tensor network: An adaptive gauge picture
  for enhanced performance},
\newblock Phys. Rev. B \textbf{90}, 125154 (2014),
\newblock \doi{10.1103/PhysRevB.90.125154}.

\bibitem{SilviCirac_binaryTree_2010}
P.~Silvi, V.~Giovannetti, S.~Montangero, M.~Rizzi, J.~I. Cirac and R.~Fazio,
\newblock \emph{Homogeneous binary trees as ground states of quantum critical
  hamiltonians},
\newblock Phys. Rev. A \textbf{81}, 062335 (2010),
\newblock \doi{10.1103/PhysRevA.81.062335}.

\bibitem{MurgVerstraete_TTN_2010}
V.~Murg, F.~Verstraete, O.~Legeza and R.~M. Noack,
\newblock \emph{Simulating strongly correlated quantum systems with tree tensor
  networks},
\newblock Phys. Rev. B \textbf{82}, 205105 (2010),
\newblock \doi{10.1103/PhysRevB.82.205105}.

\bibitem{KlaasVerstraete_TTNChemistryDMRG_2018}
K.~Gunst, F.~Verstraete, S.~Wouters, {\"O}.~Legeza and N.~D. Van,
\newblock \emph{T3ns: Three-legged tree tensor network states},
\newblock Journal of Chemical Theory and Computation \textbf{14}(4), 2026
  (2018),
\newblock \doi{https://doi.org/10.1021/acs.jctc.8b00098},
\newblock Doi: 10.1021/acs.jctc.8b00098.

\bibitem{TagliacozzoVidal_2dTTN_2009}
L.~Tagliacozzo, G.~Evenbly and G.~Vidal,
\newblock \emph{Simulation of two-dimensional quantum systems using a tree
  tensor network that exploits the entropic area law},
\newblock Phys. Rev. B \textbf{80}, 235127 (2009),
\newblock \doi{10.1103/PhysRevB.80.235127}.

\bibitem{ShiVidal_TTN_2006}
Y.-Y. Shi, L.-M. Duan and G.~Vidal,
\newblock \emph{Classical simulation of quantum many-body systems with a tree
  tensor network},
\newblock Phys. Rev. A \textbf{74}, 022320 (2006),
\newblock \doi{10.1103/PhysRevA.74.022320}.

\bibitem{FTPS}
D.~Bauernfeind, M.~Zingl, R.~Triebl, M.~Aichhorn and H.~G. Evertz,
\newblock \emph{Fork tensor-product states: Efficient multiorbital real-time
  dmft solver},
\newblock Phys. Rev. X \textbf{7}, 031013 (2017),
\newblock \doi{10.1103/PhysRevX.7.031013}.

\bibitem{THESIS}
D.~Bauernfeind,
\newblock \emph{Fork Tensor Product States: Efficient Multi-Orbital Impurity
  Solver for Dynamical Mean Field Theory},
\newblock dissertation, Graz University of Technology (2018).

\bibitem{Eisert_AreaLaw}
J.~Eisert, M.~Cramer and M.~B. Plenio,
\newblock \emph{Colloquium},
\newblock Rev. Mod. Phys. \textbf{82}, 277 (2010),
\newblock \doi{10.1103/RevModPhys.82.277}.

\bibitem{Eisert_NonEq_2015}
{Eisert J.}, {Friesdorf M.} and {Gogolin C.},
\newblock \emph{{Quantum many-body systems out of equilibrium}},
\newblock Nature Physics \textbf{11}, 124 (2015),
\newblock \doi{https://doi.org/10.1038/nphys3215 10.1038/nphys3215}.

\bibitem{Matthias_MagnetizationProfiles_2019}
M.~{Gruber} and V.~{Eisler},
\newblock \emph{{Magnetization and entanglement after a geometric quench in the
  XXZ chain}},
\newblock arXiv e-prints arXiv:1902.05834 (2019),
\newblock \eprint{1902.05834}.

\bibitem{DanielViktor_FrontDynamics}
V.~Eisler and D.~Bauernfeind,
\newblock \emph{Front dynamics and entanglement in the xxz chain with a
  gradient},
\newblock Phys. Rev. B \textbf{96}, 174301 (2017),
\newblock \doi{10.1103/PhysRevB.96.174301}.

\bibitem{Collura_domainWallexsolution_2018}
M.~Collura, A.~De~Luca and J.~Viti,
\newblock \emph{Analytic solution of the domain-wall nonequilibrium stationary
  state},
\newblock Phys. Rev. B \textbf{97}, 081111 (2018),
\newblock \doi{10.1103/PhysRevB.97.081111}.

\bibitem{DaleySchollwoek_tDMRG}
A.~J. Daley, C.~Kollath, U.~Schollwöck and G.~Vidal,
\newblock \emph{Time-dependent density-matrix renormalization-group using
  adaptive effective hilbert spaces},
\newblock Journal of Statistical Mechanics: Theory and Experiment
  \textbf{2004}(04), P04005 (2004).

\bibitem{White_tDMRG}
S.~R. White and A.~E. Feiguin,
\newblock \emph{Real-time evolution using the density matrix renormalization
  group},
\newblock Phys. Rev. Lett. \textbf{93}, 076401 (2004),
\newblock \doi{10.1103/PhysRevLett.93.076401}.

\bibitem{VidalTEBD1}
G.~Vidal,
\newblock \emph{{Efficient Classical Simulation of Slightly Entangled Quantum
  Computations}},
\newblock Phys. Rev. Lett. \textbf{91}, 147902 (2003),
\newblock \doi{10.1103/PhysRevLett.91.147902}.

\bibitem{VidalTEBD2}
G.~Vidal,
\newblock \emph{Efficient simulation of one-dimensional quantum many-body
  systems},
\newblock Phys. Rev. Lett. \textbf{93}, 040502 (2004),
\newblock \doi{10.1103/PhysRevLett.93.040502}.

\bibitem{HaegemanTDVP1}
J.~Haegeman, C.~Lubich, I.~Oseledets, B.~Vandereycken and F.~Verstraete,
\newblock \emph{{Unifying time evolution and optimization with matrix product
  states}},
\newblock Phys. Rev. B \textbf{94}, 165116 (2016),
\newblock \doi{10.1103/PhysRevB.94.165116}.

\bibitem{HaegemanTDVP2}
J.~{Haegeman}, J.~I. {Cirac}, T.~J. {Osborne}, I.~{Pi{\v z}orn},
  H.~{Verschelde} and F.~{Verstraete},
\newblock \emph{{Time-Dependent Variational Principle for Quantum Lattices}},
\newblock Phys. Rev. Lett. \textbf{107}(7), 070601 (2011),
\newblock \doi{10.1103/PhysRevLett.107.070601}.

\bibitem{Lubich_TDVP}
C.~Lubich, I.~V. Oseledets and B.~Vandereycken,
\newblock \emph{Time integration of tensor trains},
\newblock SIAM Journal on Numerical Analysis \textbf{53}(2), 917 (2015),
\newblock \doi{10.1137/140976546},
\newblock \eprint{https://doi.org/10.1137/140976546}.

\bibitem{Hubig_CompareTevoAlgs}
S.~Paeckel, T.~K{\"o}hler, A.~Swoboda, S.~R. Manmana, U.~Schollw{\"o}ck and
  C.~Hubig,
\newblock \emph{Time-evolution methods for matrix-product states},
\newblock arXiv preprint arXiv:1901.05824  (2019).

\bibitem{Rizzi_MERATevo_2008}
M.~Rizzi, S.~Montangero and G.~Vidal,
\newblock \emph{Simulation of time evolution with multiscale entanglement
  renormalization ansatz},
\newblock Phys. Rev. A \textbf{77}, 052328 (2008),
\newblock \doi{10.1103/PhysRevA.77.052328}.

\bibitem{PhienVidal_PEPSTEBD_2015}
H.~N. Phien, I.~P. McCulloch and G.~Vidal,
\newblock \emph{Fast convergence of imaginary time evolution tensor network
  algorithms by recycling the environment},
\newblock Phys. Rev. B \textbf{91}, 115137 (2015),
\newblock \doi{10.1103/PhysRevB.91.115137}.

\bibitem{Hubig_IPEPS_RealTimeTEBD}
C.~Hubig and J.~I. Cirac,
\newblock \emph{{Time-dependent study of disordered models with infinite
  projected entangled pair states}},
\newblock SciPost Phys. \textbf{6}, 31 (2019),
\newblock \doi{10.21468/SciPostPhys.6.3.031}.

\bibitem{Czarnik_IPEPSTEBD}
P.~Czarnik, J.~Dziarmaga and P.~Corboz,
\newblock \emph{Time evolution of an infinite projected entangled pair state:
  An efficient algorithm},
\newblock Phys. Rev. B \textbf{99}, 035115 (2019),
\newblock \doi{10.1103/PhysRevB.99.035115}.

\bibitem{LiXiang_BetheInfiniteDMRG_2012}
W.~Li, J.~von Delft and T.~Xiang,
\newblock \emph{Efficient simulation of infinite tree tensor network states on
  the bethe lattice},
\newblock Phys. Rev. B \textbf{86}, 195137 (2012),
\newblock \doi{10.1103/PhysRevB.86.195137}.

\bibitem{Nagaj_TEBDImagTevoTTN_2008}
D.~Nagaj, E.~Farhi, J.~Goldstone, P.~Shor and I.~Sylvester,
\newblock \emph{Quantum transverse-field ising model on an infinite tree from
  matrix product states},
\newblock Phys. Rev. B \textbf{77}, 214431 (2008),
\newblock \doi{10.1103/PhysRevB.77.214431}.

\bibitem{SuzukiDecomp}
M.~Suzuki,
\newblock \emph{{Fractal decomposition of exponential operators with
  applications to many-body theories and Monte Carlo simulations}},
\newblock Physics Letters A \textbf{146}(6), 319  (1990).

\bibitem{Meyer_MCHF1}
H.-D. Meyer, U.~Manthe and L.~Cederbaum,
\newblock \emph{The multi-configurational time-dependent hartree approach},
\newblock Chemical Physics Letters \textbf{165}(1), 73  (1990),
\newblock \doi{https://doi.org/10.1016/0009-2614(90)87014-I}.

\bibitem{Meyer_MCHF2}
U.~Manthe, H.~Meyer and L.~S. Cederbaum,
\newblock \emph{Wave‐packet dynamics within the multiconfiguration hartree
  framework: General aspects and application to nocl},
\newblock The Journal of Chemical Physics \textbf{97}(5), 3199 (1992),
\newblock \doi{10.1063/1.463007},
\newblock \eprint{https://doi.org/10.1063/1.463007}.

\bibitem{Manthe_MCHF}
U.~Manthe,
\newblock \emph{A multilayer multiconfigurational time-dependent hartree
  approach for quantum dynamics on general potential energy surfaces},
\newblock The Journal of Chemical Physics \textbf{128}(16), 164116 (2008),
\newblock \doi{10.1063/1.2902982},
\newblock \eprint{https://doi.org/10.1063/1.2902982}.

\bibitem{Manthe_MCHF2}
U.~Manthe,
\newblock \emph{Wavepacket dynamics and the multi-configurational
  time-dependent hartree approach},
\newblock Journal of Physics: Condensed Matter \textbf{29}(25), 253001 (2017),
\newblock \doi{10.1088/1361-648x/aa6e96}.

\bibitem{Zaletel_longrange}
M.~P. Zaletel, R.~S.~K. Mong, C.~Karrasch, J.~E. Moore and F.~Pollmann,
\newblock \emph{Time-evolving a matrix product state with long-ranged
  interactions},
\newblock Phys. Rev. B \textbf{91}, 165112 (2015),
\newblock \doi{10.1103/PhysRevB.91.165112}.

\bibitem{RamsEntBarrier}
M.~M. {Rams} and M.~{Zwolak},
\newblock \emph{{Breaking the entanglement barrier: Tensor network simulation
  of quantum transport}},
\newblock arXiv e-prints arXiv:1904.12793 (2019),
\newblock \eprint{1904.12793}.

\bibitem{schroederF_TDVP}
F.~A. Y.~N. Schr\"oder and A.~W. Chin,
\newblock \emph{Simulating open quantum dynamics with time-dependent
  variational matrix product states: Towards microscopic correlation of
  environment dynamics and reduced system evolution},
\newblock Phys. Rev. B \textbf{93}, 075105 (2016),
\newblock \doi{10.1103/PhysRevB.93.075105}.

\bibitem{Lubich_TTNTDVP}
C.~Lubich, T.~Rohwedder, R.~Schneider and B.~Vandereycken,
\newblock \emph{Dynamical approximation by hierarchical tucker and tensor-train
  tensors},
\newblock SIAM Journal on Matrix Analysis and Applications \textbf{34}(2), 470
  (2013),
\newblock \doi{10.1137/120885723},
\newblock \eprint{https://doi.org/10.1137/120885723}.

\bibitem{Silvi_TensorNetworksAnthology_2019}
P.~Silvi, F.~Tschirsich, M.~Gerster, J.~Jünemann, D.~Jaschke, M.~Rizzi and
  S.~Montangero,
\newblock \emph{{The Tensor Networks Anthology: Simulation techniques for
  many-body quantum lattice systems}},
\newblock SciPost Phys. Lect. Notes p.~8 (2019),
\newblock \doi{10.21468/SciPostPhysLectNotes.8}.

\bibitem{Kanamori}
J.~Kanamori,
\newblock \emph{Electron correlation and ferromagnetism of transition metals},
\newblock Progress of Theoretical Physics \textbf{30}(3), 275 (1963),
\newblock \doi{10.1143/PTP.30.275}.

\bibitem{StarVSChainsPaper}
D.~{Bauernfeind}, M.~{Aichhorn} and H.~G. {Evertz},
\newblock \emph{{Comparison of MPS based real time evolution algorithms for
  Anderson Impurity Models}},
\newblock arXiv e-prints arXiv:1906.09077 (2019),
\newblock \eprint{1906.09077}.

\bibitem{Kohn_TTNTDVP}
L.~{Kohn}, P.~{Silvi}, M.~{Gerster}, M.~{Keck}, R.~{Fazio}, G.~E. {Santoro} and
  S.~{Montangero},
\newblock \emph{{Superfluid to Mott transition in a Bose-Hubbard ring:
  Persistent currents and defect formation}},
\newblock arXiv e-prints arXiv:1907.00009 (2019),
\newblock \eprint{1907.00009}.

\end{thebibliography}

\end{document}